\documentstyle[12pt,aasms4]{article}
\begin{document}

\title{A Revised Ephemeris and FUSE Observations \\ of the Supersoft X-ray
Source CAL~83\footnote{This paper utilizes public domain data obtained by
the MACHO Project, jointly funded by the US Department of Energy through
the University of California, Lawrence Livermore National Laboratory under
contract No.\ W-7405-Eng-48, by the National Science Foundation through
the Center for Particle Astrophysics of the University of California under
coopertative agreement AST-8809616, and by the Mount Stromlo and Siding
Spring Observatory, part of the Australian National University. }
\footnote{Based on observations made with the NASA-CNES-CSA Far
Ultraviolet Spectroscopic Explorer.  FUSE is operated for NASA by the
Johns Hopkins University under NASA contract NAS5-3298.} } 

\author{P.C. Schmidtke\altaffilmark{3}, A.P. Cowley}  
\affil{Department of Physics \& Astronomy, Arizona State University,
Tempe, AZ, 85287-1504 \\
email: paul.schmidtke@asu.edu}

\and 

\author{J.B. Hutchings, K. Winter, \& D. Crampton}
\affil{Herzberg Institute of Astrophysics, National Research Council of
Canada, \\ 5071 W.\ Saanich Rd., Victoria, B.C. V9E 2E7, Canada }

\altaffiltext{3} {Visiting Astronomer, Cerro Tololo Inter-American
Observatory, National Optical Astronomy Observatories, which is operated
by the Association of Universities for Research in Astronomy, Inc., under
contract with the National Science Foundation.}

\begin{abstract}

A new ephemeris has been determined for the supersoft X-ray binary CAL~83
using MACHO photometry.  With an improved orbital period of 1.047568 days,
it is now possible to phase together photometric and spectroscopic data
obtained over the past two decades with new far ultraviolet spectra taken
with FUSE.  We discuss the properties of the orbital and longterm optical
light curves as well as the colors of CAL~83.  In the far ultraviolet the
only well-detected stellar feature is emission from the O~VI resonance
doublet.  The radial velocity of this emission appears to differ from that
of He~II in the optical region, although we only have partial phase
coverage for the O~VI line.  The FUSE continuum variations are similar to
the optical light curve in phase and amplitude. 

\end{abstract}


\section{Introduction}

The X-ray binary CAL~83 is one of a small number of supersoft sources
which are characterized by an extremely soft X-ray spectrum with almost no
radiation above $\sim0.5$ keV.  This group includes a handful of sources
in the Magellanic Clouds and the Galaxy (e.g. Greiner 1996, Hasinger 1996,
Cowley et al. 1998), plus some less well studied objects in other nearby
galaxies (e.g. Kahabka 1999; Swartz et al. 2002; Kong \& Di Stefano 2003).
CAL~83 itself is a member of the LMC, based on its position and radial
velocity. 

From radial velocity variations of the optical emission lines (especially
He~II and H), Crampton et al.\ (1987) discovered that CAL~83 is a binary
system with a period near one day.  The emission lines are formed in an
accretion disk that is viewed at a low inclination angle, so the resulting
velocity amplitude is small (K$\sim30$ km s$^{-1}$).  There is no
spectroscopic evidence of either star, but the period and velocities
suggest that they have masses well under 2M${\odot}$ (see Crampton et al.
1987, Smale et al. 1988). 

CAL~83 is known to exhibit irregular brightness variations of more than a
magnitude over timescales of months (e.g. Smale et al. 1988, Alcock et al.
1997, Greiner \& Di Stefano 2002).  Although CAL~83 is considered a nearly
steady X-ray source, occasionally it drops into an X-ray-off state (e.g.
Kahabka 1996, 1998).  Using MACHO data, Alcock et al.\ and Greiner \&
Di Stefano have shown that these low X-ray states are followed within
$\sim30-50$ days by a low optical state.  The most likely explanation is
that there is a change in the accretion flow and hence in the accretion
disk which provides the optical luminosity. 

In spite of these large changes in the mean brightness of CAL~83, Smale et
al.\ were able to detect small superimposed orbital variations with a full
amplitude of $\sim$0.2 mag and a period of P = 1.0436 days.  During the
orbital period the light curve shows a sinusoidal variation with one
maximum and one minimum per cycle.  Cowley et al.\ (1991) later refined
the orbital period to be P = 1.0475 days by combining all published and
new spectroscopic data as well as the photometry of Smale et al. 

Given the low inclination of the system, it is unlikely that the minimum
is caused by an occulatation by the secondary star.  Furthermore, the
phasing of the emission line velocities implies that the light minimum
does not coincide with conjunction of the stars.  The variation is likely
to arise from a partial obscuration of the bright inner accretion disk. 
Possible occulting regions might be a splash region raised by impact of
the accretion stream or a gas stream which is significantly above the
orbital plane.  Although we are uncertain what causes the orbital light
curve, we have been able to use it to refine the orbital period to allow
intercomparison of two decades of spectroscopic data. 

\section{Analysis of Optical Data}

\subsection{Calibration of the MACHO data}

CAL~83 was observed for $\sim$7 years by the MACHO Project.  We extracted
both the ``blue" and ``red" data for this star using the interactive
online web browser\footnote{See
http://www.macho.mcmaster.ca/Data/MachoData.html}.  After transformation of
the instrumental magnitudes, using the method described by Alcock et al.\
(1999), the derived values are close to $V$ and $R$ band magnitudes.  We
refer to these as $V_{MAC}$ and $R_{MAC}$, as we have done in previous
papers.  The linear transformation equations adopted here differ slightly
from those given by Greiner \& Di Stefano (2002), in the sense that our
$V_{MAC}$ and $R_{MAC}$ values are 0.012 mag brighter and 0.013 mag
fainter, respectively, than their ``generic" photometry.  The combination
of offsets makes our $V_{MAC}-R_{MAC}$ colors 0.025 magnitudes bluer.  All
of these systematic differences, however, are small compared to the
estimated zero-point calibration accuracies ($\pm$0.035 mag) given by
Alcock et al.  Our primary concern is with the internal errors of the
MACHO data.  Hence we propogate, in all of our calculations, only the
photometric uncertainties of the instrumental magnitudes.  In applying the
transformations it was assumed that the $V_{MAC}-R_{MAC}$ color for CAL~83
is constant, but small variations are present in the MACHO data (see
$\S$2.2). 

Figure 1 shows the $V_{MAC}$ light curve for CAL~83.  The source is highly
variable with rapid changes in brightness between high, intermediate, and
low optical states (see Greiner \& Di Stefano).  There were two epochs
when the mean magnitude remained relatively stable (MJD 48917--49640 and
50980--51541).  CAL~83 was in an intermediate state during Segment 1 and a
high state during Segment 2. 

\subsection{Color Variations Based on MACHO Data}

In order to understand what causes the light variations, it is of interest
to determine whether or not the color changes as the system brightens.
Greiner \& Di Stefano (2002), in their study of the long-term on/off
states, found no relationship between color and brightness.  They pointed
out that the reddest $V-R$ data have very large errors and argued that the
mean $V-R$ color varies by no more than 0.03$\pm$0.05 mag for $>$1 mag
changes in system brightness (see their Fig.\ 2).  However, we find that
the mean $V_{MAC}-R_{MAC}$ depends on CAL~83's optical state.  In Figure 2
we plot $V_{MAC}$ as a function of $V_{MAC}-R_{MAC}$.  The dividing lines
between high, intermediate, and low states roughly correspond to the
limits set by Greiner \& Di Stefano.  One sees that the color is bluest
when the system is in its high state, although the change is small. 

Removing the 17 points in the figure with the largest uncertainties (i.e.,
error estimates $\ge$0.048 mag), the remaining points have a {\it
distribution of errors} that approximates the expectation for a Gaussian
distribution with $\sigma \sim 0.018$ mag.  That is, $\sim$68\% of the
remaining points have errors $\le$1$\sigma$ and $\sim$97\% have errors
$\le$2$\sigma$.  This close agreement with the expected percentages for a
normal distribution (i.e., 68\%, 95\%, etc.) validates our selection of
points to remove from the analysis.  We find the mean $V_{MAC}-R_{MAC}$
colors are $-0.065{\pm}0.019$ (56 points), $0.000{\pm}0.038$ (66 points),
and $-0.024{\pm}0.027$ (31 points) for the high, intermediate, and low
states, respectively.  The observed dispersion within each sample is
comparable to the internal precision of the entire MACHO dataset
($\sigma_{V-R}=0.028$ mag for $V \le 18$ mag) as demonstrated by Alcock
et al. (1999).  Simple T-tests show that the samples are significantly
different compared to each other.  However, systematic errors are present in
the means because we have assumed constant color terms in the transformation
equations.  Hence, we have slightly overestimated the color differences
between states.  The largest mean difference (between high and intermediate
states) is $-$0.065 mag, but after compensating for the transformation color
terms, the revised differential is ${\sim}-$0.048 mag.  The corrected value
is still much greater than the expected internal precision of MACHO data
(i.e., 0.028 mag) and therefore judged to a meaningful difference.  All of
the other corrections are smaller than this example.  After making the
adjustments, the color difference of even the smallest differential (between
intermediate and low states) is still significant at a 98\% level. 

\subsection{Determination of a Revised Orbital Ephemeris from MACHO Data}

After removal of a few errant (i.e., spuriously faint) points, data from
the two time segments marked in Figure 1 were searched for orbital
variations.  Since the data are more complete for $V_{MAC}$, observations
in only that bandpass were used to redetermine the orbital period.  The
$V_{MAC}$ magnitudes for each segment were analyzed separately to derive
periodograms, using the technique described by Horne \& Baliunas (1986). 
Both segments contain one significant peak with very similar values near P
= 1.04 days.  (Due to the daily sampling window, several aliases also are
present.)  Therefore, we combined the data from the two segments after
subtracting the mean linear fit to each segment.  This resulted in a
single set of relative magnitudes, $V^*$, from which the combined power
spectrum, shown in Figure 3, was determined.  Fitting a sinewave to the
$V^*$ data yielded the following ephemeris: 

\centerline{T(min) = JD~2451500.953$\pm$0.004 + 1.047568$\pm$0.000003E 
days} 

\noindent
For the epoch of the FUSE observations presented in this paper (see $\S$3),
the new ephemeris gives phases 0.15 earlier than those predicted by Smale
et al.\ (1988), so this revision of the orbital period is important for 
correctly phasing the FUSE data. 

Plots of the $V^*$ and $R^*$ light curves, folded on the new ephemeris,
are shown in Figure 4, with points from the two time segments indicated by
different symbols.  We note that the amplitude of the orbital variation
appears to depend on the mean brightness level in both bandpasses.  The
$V_{MAC}$ data show that when the system was at $V_{MAC}$ = 17.34 during
the first segment, the semiamplitude of the variation was 0.053$\pm$0.001
mag, but during the second segment (mean magnitude $V_{MAC}$ = 16.97) the
semiamplitude was 0.075$\pm$0.001 mag.  Similarly, the $R_{MAC}$ data show
that the amplitude in the first (fainter) time segment was
0.050$\pm$0.003, while in the last (brighter) segment the amplitude was
0.082$\pm$0.002.  In addition, the data of Smale et al.\ have a
semiamplitude of 0.11 mag when the system had a mean magnitude of $V$ =
16.87.  Hence, the amplitude of the orbital modulation appears to increase
as the system brightens.
  
Smale et al.\ (1988) do not tabulate their photometric data, so we are
unable to plot their data on the MACHO ephemeris.  However, using Smale et
al.'s ephemeris, we estimate minimum light occurred near HJD 2446051.585,
which corresponds to our phase 0.07.  The discrepancy is small but
reasonable, given that their assumed time of minimum is based on an analysis
of data which are folded on a somewhat different orbital period and may
contain fluctuations in mean brightness. 

In the bottom panel of Figure 4 we plot $V_{MAC}-R_{MAC}$ versus orbital
phase.  Although there is no significant variation of color with phase, the
figure confirms that the mean color in the second segment (when the source
was in a high optical state) is bluer than in the first segment
(intermediate state), as found in $\S$2.2. 

\subsection{Previously Unpublished Photometry of CAL~83}

$B$ and $V$ photometry was obtained on 8 nights in 1999 March using the
TEK2K \#3 CCD on the CTIO 0.9 m telescope.  The images were reduced with
DAOPHOT (Stetson 1987) and calibrated via observations of Landolt's (1992)
standard stars.  Differential photometry was then calculated using a set of
reference stars in the field of view of CAL~83.  The same comparison stars
were used in our previous photometric studies (Crampton et al. 1987,
Cowley et al. 1998), hence all of our measurements are on a common scale. 
The new photometric measures are listed in Table 1 and shown
in Figure 5.  For a short observing run, the $\sim$1 day orbital period
results in poor phase coverage, so it is not possible to use these data to
determine an independent value for the period.  Hence, the $V$ light curve
in the top panel of the figure is fitted using a sinewave with P and
T(min) fixed at the new values derived in $\S$2.3.  The resulting
semiamplitude (0.062$\pm$0.012 mag) and mean magnitude ($V$ =
16.986$\pm$0.009) are consistent with the MACHO reductions for the same
epoch.  The color, plotted in the bottom panel, shows little variation and
has a mean value of $B-V$ = $-$0.059$\pm$0.015. 

\section{Re-analysis of Published Spectroscopic Data Using the New Ephemeris}

We have gone back to the spectroscopic data given by both Crampton et al.\
(1987) and by Smale et al.\ (1988).  Using the He~II (4686\AA) velocities
from Crampton et al., we have computed revised orbital elements for CAL~83
which are presented in Table 2.  A plot of the He~II velocities (from
Crampton et al.) versus phase is shown in the lower panel of Figure 6,
where it is compared to the FUSE O~VI data.  Although the He~II velocities
tabulated by Smale et al.\ are not plotted, their phasing and amplitude
are very similar.  We note that their velocities appear to be
systematically higher by $\sim$30 km s$^{-1}$.  It is likely that this is
related to the longterm variations in the shape of the He~II line pointed
out by Crampton et al.  Over a timescale of months, the $\lambda$4686 line
shows wings extending several thousand km s$^{-1}$, alternately to longer
and then shorter wavelengths.  Thus, depending on the overall line shape
during the epoch when observations were made, the mean He~II wavelength
might be measured somewhat differently, thus accounting for apparent
shifts in the systemic velocity. 

From HST data observed 1996 November 10 (our orbital phase $\Phi$ = 0.69),
G\"ansicke et al.\ (1998) measured a velocity of +300 km s$^{-1}$ for the
N~V doublet at $\lambda\lambda$1239 \& 1243.  This value consistent with
our He~II velocity curve.  They note that the lines are asymmetrical with
a redward wing extending to $\sim$+800 km s$^{-1}$. 

The emission lines are typical of those from an accretion disk which
surrounds the compact star.  Provided there is no strong contribution to
these lines from the mass transfer stream, then they should trace the
orbital motion of this star.  The velocity curve shows that minimum light
occurs about 0.12P before the compact star is at superior conjunction. 

Data in the literature show that the mean He~II equivalent widths (EW)
changes over the long term, and probably with phase, although this
information is incomplete for most epochs.  Our 1984 January spectra from
Las Campanas have a mean EW = $-5$\AA, while spectra from 1985 December
when the system was faint ($V\sim17.3$, Crampton et al. 1987) show nearly
double that value.  This suggests that when the system is faint the
emission line EW may be greater, but further data are needed to confirm
this. 

\section{FUSE Observations and Measurements}

In 2001 September, CAL~83 was observed using FUSE.  This instrument is
sensitive over the wavelength range 912--1187\AA, at a spectral resolution
of R = 20000 or more.  FUSE has four independent optical channels, feeding
two detectors via Rowland circle spectrographs.  Two optical trains use
LiF-coated optics (990--1187\AA) and two have SiC-coated optics for
wavelengths down to 912\AA.  As each detector is in two sections, there
are 8 channels of data, designated Lif1a, Lif1b, etc., through Sic2b.  A
more complete description of the instrument is given by Moos et al.\ (2000).
  
The observations of CAL~83 were taken during 11 consecutive FUSE orbits,
with earth-occultation interruptions.  The star was located in the large
aperture, and the alignment of the FUSE channels should have been stable
over the whole observation.  The visibility times were generally of order
4000 ksec.  The whole observing program covered about 0.6P of the binary
orbit.  The basic observational details are given in Table 3.  The data
were extracted and processed with CALFUSE, current in April 2002.  Because
the source is faint, the airglow spectrum constitutes a major contaminant
and effectively obliterates a significant part of the FUSE spectral range.

The only stellar line feature seen in the FUSE spectrum is O~VI emission
at 1031.9\AA.  The second line of the O~VI doublet, 1038\AA, is overlaid
by airglow lines with peak emission some ten times that of the O~VI line. 
The 1032\AA\ line is weak, so in order to measure the continuum and O~VI
with more reliability we have binned the observations into 5 spectra,
using consecutive pairs except for the last 3 observations which were
summed to form a single spectrum.  Figure 7 shows the binned spectra in
the region of the 1032\AA\ O~VI emission line, together with some nearby
absorption features.  These spectra combined the LiF channels only, as the
SiC channels were very noisy. 

The continuum flux was measured in the airglow-free wavelength region
1045-1080\AA, in both the LiF1a and LiF2b channels.  Similar changes were
seen in both, but the LiF2b showed a much greater range, consistent with a
poorly-determined zero level in the extraction.  As the LiF1a channel is
that used for guiding, we have adoped continuum values from this channel
only.  (The profiles shown in Figure 7 include the LiF2b to maximize the
signal, but the O~VI profile is not affected by zero point uncertainties.)
The continuum flux values, given in Table 3, have a mean value of
$\sim1.2\times10^{-14}$ erg s$^{-1}$ cm$^{-2}$ \AA$^{-1}$ and are almost
flat across the FUSE bandpass.  Since the changes in the continuum are
similar for the LiF1a and LiF2b channels, we consider them to be real and
note that they have a range and phasing similar to the MACHO data.  Figure
6 shows the FUSE continuum changes compared with the mean optical light
curve. 

The O~VI line was measured for flux and radial velocity, and these values
are also given in Table 3.  We estimate the uncertainty in measurement of
the line flux is no larger than $\sim$20\% of its value.  The O~VI flux
appears to be greatest between phases 0.5-0.8 and weakest near phase 0.1,
although the phase coverage is incomplete.  The O~VI flux variation
appears to be similar to that of the optical He~II line, for the
overlapping phases.  Additional FUSE observations are needed to see if the
observed O~VI flux variations are phase related. 

The O~VI line was measured in several ways, first by profile fitting of
line and then by using the centroid of the flux above the local continuum.
In addition, the velocities were measured by cross-correlation against the
mean CAL~83 spectrum.  There is close agreement (in all cases $<$15 km
s$^{-1}$) between the profile fitting technique and the cross-correlation
velocities.  The adopted values given in Table 3 are the mean of the two
methods.  We estimate the error of each measure to be no more than $\pm$20
km s$^{-1}$.  Both sets of measured velocities are plotted versus phase in
the bottom panel of Figure 6, with different symbols for each method to
show the close agreement.  In this figure we also show the He~II
velocities from Crampton et al.\ (1987). 

As a check on the stability of the wavelength scale, the double absorption
feature of C~II at 1036--7\AA\ was also measured by cross-correlation of
only this part of the spectrum against the mean for all CAL~83 FUSE
spectra.  Although the C~II lines are weak and hence difficult to measure,
their velocities show no correlation with each other or with the O~VI
velocities.  Even though in Figure 6 it may appear that their positions
are not constant, noise in these very weak features makes the profiles
and/or the doublet separation appear to change.  The measurements of C~II
are consistent with no velocity variation, with a scatter of about $\pm$12
km s$^{-1}$. 

Finally, the wavelength scale itself was checked by measuring the airglow
emissions near 1038\AA.  These airglow lines were found to have the same
wavelength for all binned spectra, to within $\pm$0.01\AA, but their
average position is too high by 15 km s$^{-1}$.  This wavelength scale
error is typical of large aperture observations, but may partly reflect
the average FUSE orbital motion through the visibility windows to the
target.  We have applied a correction of +15 km s$^{-1}$ to the O~VI
velocity measures. 

We note that the mean O~VI velocity of +434 km s$^{-1}$ is much more
positive than the systemic velocity of +257 km s$^{-1}$ given here from
He~II, 4686\AA\ emission.  It is possible that the shortward edge of the
line has P~Cygni-like absorption (see Figure 7), which may account for
some of the velocity offset.  However, the velocity difference is more
than 0.5\AA, so this explanation would imply a much broader and stronger
intrinsic emission profile and a large absorption in a disk wind.  As
noted above, the longterm changes in the emission wings of He~II were
reported by Crampton et al.\ (1987; also see discussion in Cowley et al.
1998).  It is possible that the offset seen in the velocities from the
FUSE data may be partially due to the long wavelength extension of the
line profile, which can be seen in Figure 7. 

Fitting the O~VI FUSE velocities to a sinewave results in a semiamplitude
K = 23$\pm$6 km s$^{-1}$ with maximum velocity occuring at photometric
phase $\Phi = 0.23\pm0.03$.  Thus, the phasing of the O~VI emission
appears to differ from He~II, whose maximum velocity is at $\Phi = 0.88$
with a semiamplitude of K = 27 km s$^{-1}$ (see Table 2).  If further
observations in the far ultraviolet confirm this difference, then the O~VI
emission may arise in a different location from the He~II. 

\section{Summary}

To summarize, we have used MACHO photometry to refine the orbital period
of the LMC supersoft X-ray source CAL~83.  With our new ephemeris, we have
been able to compare previously published spectroscopic and photometric
data using consistent phases.  In a high optical state, its color becomes
bluer and the amplitude of the orbital variations increases.  New FUSE
observations show that orbital variations in brightness in the far
ultraviolet are similar to those in the optical region.  However, the
velocity of the O~VI emission in the FUSE spectra indicates its behavior
may differ from the optical He~II line. 

\acknowledgments 

We acknowledge help from the CTIO staff when our photometric observations
were made.

\clearpage

\begin{deluxetable}{ccc}
\tablenum{1}
\tablecaption{1999 March $B$ and $V$ Photometry of CAL~83 }
\tablehead{
\colhead{HJD} &
\colhead{$B$} &
\colhead{$V$} \\
\colhead{2450000+}
}
\startdata
1258.5091 & ... & 16.923$\pm$0.021 \nl
1258.5138 & 16.863$\pm$0.017 & ... \nl
1258.6141 & ... & 16.967$\pm$0.017 \nl
1258.6187 & 16.910$\pm$0.015 & ... \nl
1259.5527 & ... & 16.994$\pm$0.013 \nl
1259.5585 & 16.914$\pm$0.010 & ... \nl
1259.5801 & ... & 16.965$\pm$0.017 \nl
1259.5844 & 16.905$\pm$0.052 & ... \nl
1259.5995 & ... & 16.987$\pm$0.019 \nl
1259.6038 & 16.912$\pm$0.017 & ... \nl
1260.5576 & ... & 16.899$\pm$0.011 \nl
1260.5629 & 16.824$\pm$0.010 & ... \nl
1260.5793 & ... & 16.909$\pm$0.018 \nl
1260.5836 & 16.846$\pm$0.011 & ... \nl
1260.5979 & ... & 16.900$\pm$0.011 \nl
1260.6020 & 16.860$\pm$0.011 & ... \nl
1261.5575 & ... & 16.936$\pm$0.020 \nl
1261.5619 & 16.883$\pm$0.025 & ... \nl
1261.5778 & ... & 16.935$\pm$0.019 \nl
1261.5821 & 16.884$\pm$0.013 & ... \nl
1261.5965 & ... & 16.947$\pm$0.010 \nl
1261.6007 & 16.888$\pm$0.018 & ... \nl
1262.4933 & ... & 16.927$\pm$0.013 \nl
1262.4980 & 16.845$\pm$0.021 & ... \nl
1262.5427 & ... & 16.919$\pm$0.014 \nl
1262.5479 & 16.867$\pm$0.007 & ... \nl
1262.5641 & ... & 16.932$\pm$0.014 \nl
1262.5686 & 16.885$\pm$0.024 & ... \nl
1262.5824 & ... & 16.960$\pm$0.011 \nl
1262.5866 & 16.893$\pm$0.025 & ... \nl
1263.5306 & ... & 16.899$\pm$0.016 \nl
1263.5354 & 16.844$\pm$0.023 & ... \nl
1263.5526 & ... & 16.894$\pm$0.010 \nl
1263.5571 & 16.828$\pm$0.022 & ... \nl
1263.5723 & ... & 16.890$\pm$0.021 \nl
1263.5768 & 16.837$\pm$0.011 & ... \nl
1264.5220 & ... & 16.927$\pm$0.011 \nl
1264.5267 & 16.860$\pm$0.026 & ... \nl
1264.5417 & ... & 16.945$\pm$0.010 \nl
1264.5464 & 16.874$\pm$0.021 & ... \nl
1265.5330 & ... & 17.016$\pm$0.015 \nl
1265.5377 & 16.974$\pm$0.021 & ... \nl
1265.5530 & ... & 17.002$\pm$0.011 \nl
1265.5578 & 16.983$\pm$0.014 & ... \nl
\enddata
\end{deluxetable}

\clearpage

\begin{deluxetable}{ll}
\tablenum{2}
\tablecaption{Revised Orbital Elements for CAL~83 from He~II 4686\AA}
\tablehead{
\colhead{Parameter} &
\colhead{Revised Value} \\
}
\startdata
Time of maximum velocity  & T(max.\ vel.) = JD 2446412.785$\pm$0.008 \nl
Photometric phase of max.\ vel. & $\Phi$(max.\ vel.) = 0.875$\pm$0.008  \nl
Velocity semiamplitude & K = 27.0$\pm$1.6 km s$^{-1}$ \nl
Systemic velocity & V$_0$ = +257.0$\pm$1.0 km s$^{-1}$ \nl
Orbital period & P = 1.047568$\pm$0.000003 days (fixed) \nl
Orbital eccentricity & e = 0 (assumed) \nl
\enddata
\end{deluxetable}

\clearpage

\begin{deluxetable}{ccccccc}
\tablenum{3}
\tablecaption{FUSE Observations and Data for CAL~83}
\tablehead{
\colhead{MJD} &
\colhead{Exp time} &
\colhead{\% Obs} &
\colhead{$\Phi_{phot}$\tablenotemark{a}} &
\multicolumn{3}{c}{Measurements from Binned FUSE Spectra\tablenotemark{b}} \\
\colhead{(mid-exp)} &
\colhead{(sec)} &
\colhead{ at night} &&
\multicolumn{2}{c}{O~VI Emission} &
\colhead{Continuum} \\
&&&& \colhead{Velocity} &
\colhead{Flux} &
\colhead{Flux\tablenotemark{c}} \\
&&&& \colhead{(km s$^{-1}$)} &
\colhead{(erg s$^{-1}$ cm$^{-2}$)} &
\colhead{(erg s$^{-1}$ cm$^{-2}$ \AA$^{-1}$) } 
}
\startdata
52174.550 & 4110 & 60 & 0.49 &  \nl
52174.620 & 4320 & 40 & 0.55 &  +430 & $8.7\times10^{-15}$ 
 & $12.6\times10^{-15}$ \nl
52174.693 & 4270 & 40 & 0.62 &  \nl
52174.766 & 4065 & 50 & 0.69 & +430 & $7.4\times10^{-15}$ 
 & $12.7\times10^{-15}$ \nl
52174.838 & 4005 & 45 & 0.76 & \nl
52174.912 & 4032 & 30 & 0.83 & +409 & $8.7\times10^{-15}$ 
 & $12.5\times10^{-15}$ \nl
52174.985 & 4302 & 25 & 0.90 &  \nl
52175.058 & 4548 & 20 & 0.97 & +440 & $6.7\times10^{-15}$ 
 & $11.0\times10^{-15}$ \nl
52175.132 & 4955 & 25 & 0.04 & \nl
52175.182 & 2973 & 50 & 0.09 &  \nl
52175.217 & 2980 & 20 & 0.12 & +453 & $5.9\times10^{-15}$ 
 & $11.4\times10^{-15}$ \nl
\enddata

\tablenotetext{a}{ Photometric ephemeris: MJD 51500.453$\pm$0.004 + 
1.047568$\pm$0.000003E, where $\Phi$ = 0 is minimum light. }

\tablenotetext{b}{ Spectra were added in pairs, except for 
the last three which were combined. }

\tablenotetext{c}{ at $\sim$1060\AA\ } 

\end{deluxetable}

\clearpage

\begin{figure}
\caption{Longterm MACHO $V_{MAC}$ light curve of CAL~83 covering dates
from 1993 January through 1999 December.  The two approximately flat
segments at the beginning and end of the observations were used in the
period analysis. } 
\end{figure}

\begin{figure}
\caption{$V_{MAC}-R_{MAC}$ color of CAL~83 as a function of $V_{MAC}$. 
Dashed lines separate the the high, intermediate, and low optical states. 
The system is bluest when it is in the high state although the change of
color is small.  Open symbols show data points with large errors which have
not been used in the analysis. } 
\end{figure}

\begin{figure}
\caption{Periodograms for CAL~83 using MACHO $V^*$ data from MJD 48917 to
49640 and from MJD 50980 to 51541.  {\it Top:} A wide range of test
frequencies is shown with the orbital periodicity and various aliases
identified.  {\it Bottom:} A narrow range of frequencies is plotted to
show details near the orbital period. } 
\end{figure}

\begin{figure}
\caption{$V^*$ ({\it top}) and $R^*$ ({\it middle}) light curves of CAL~83
plotted on P = 1.047568 days and T(min) = JD 2451500.953 (MJD 51500.453).
Different symbols show the two epochs of MACHO data that were used.
Crosses are from the first segment of MACHO data (MJD 48917 to 49640) and
open circles are from the second segment (MJD 50980 to 51541).  The fitted
sinewaves (solid lines for Segment 1 and dashed lines for Segment 2) show
the semiamplitude is larger when the source is in its high optical state.
{\it Bottom:} $V_{MAC}-R_{MAC}$ versus orbital phase is plotted using the
same symbols.  Note that the system is bluer in the second segment data,
although there is no apparent color trend with orbital phase. } 
\end{figure}

\begin{figure}
\caption{CTIO $V$ and $B-V$ data for CAL~83 from 1999 March.  The fitted
sinewave is based on the revised ephemeris, with P and T(min) fixed at
their MACHO values.  The CTIO data overlap a small part of the MACHO data
when the source was at a bright plateau. } 
\end{figure}

\begin{figure}
\caption{Comparison of FUSE and optical data.  {\it Top:} FUSE continuum
measures superimposed with a schematic $V^*$ light curve.  {\it Bottom:}
O~VI, 1032\AA, velocity curve compared to the He~II 4686\AA\ velocity
curve.  The different symbols for O~VI represent measurements using
profile fitting and cross-correlation against the mean spectrum, as
described in the text.} 
\end{figure}

\begin{figure}
\caption{Binned FUSE spectra of CAL~83 showing the region around the
1032\AA\ line of O~VI.  The mean wavelength of the O~VI line is shown as a
dotted vertical line, and the rest positions of O~VI and C~II are marked
to identify the absorptions.  The rise to longer wavelengths at the end of
the plot is due to airglow emission.  } 
\end{figure}

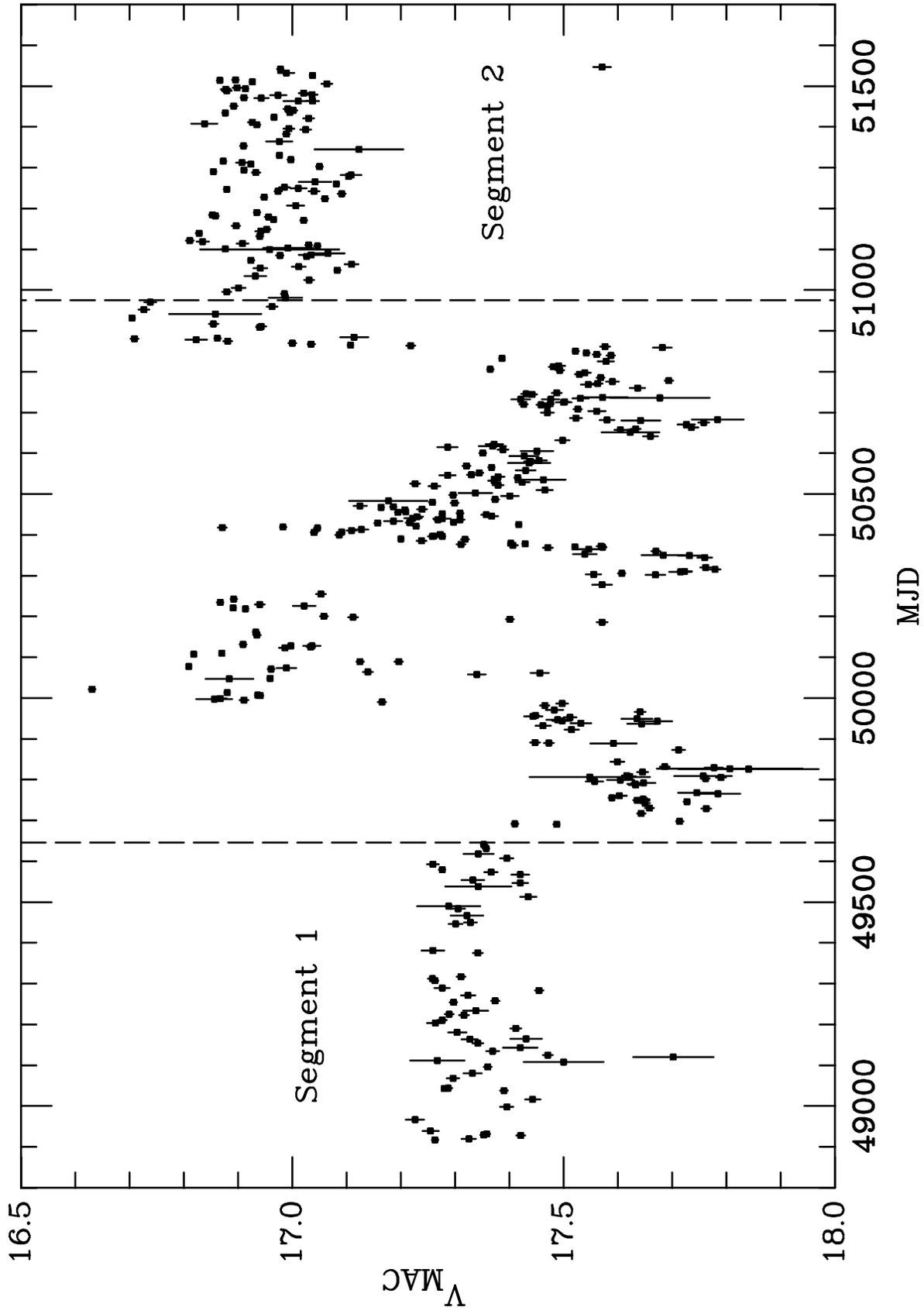
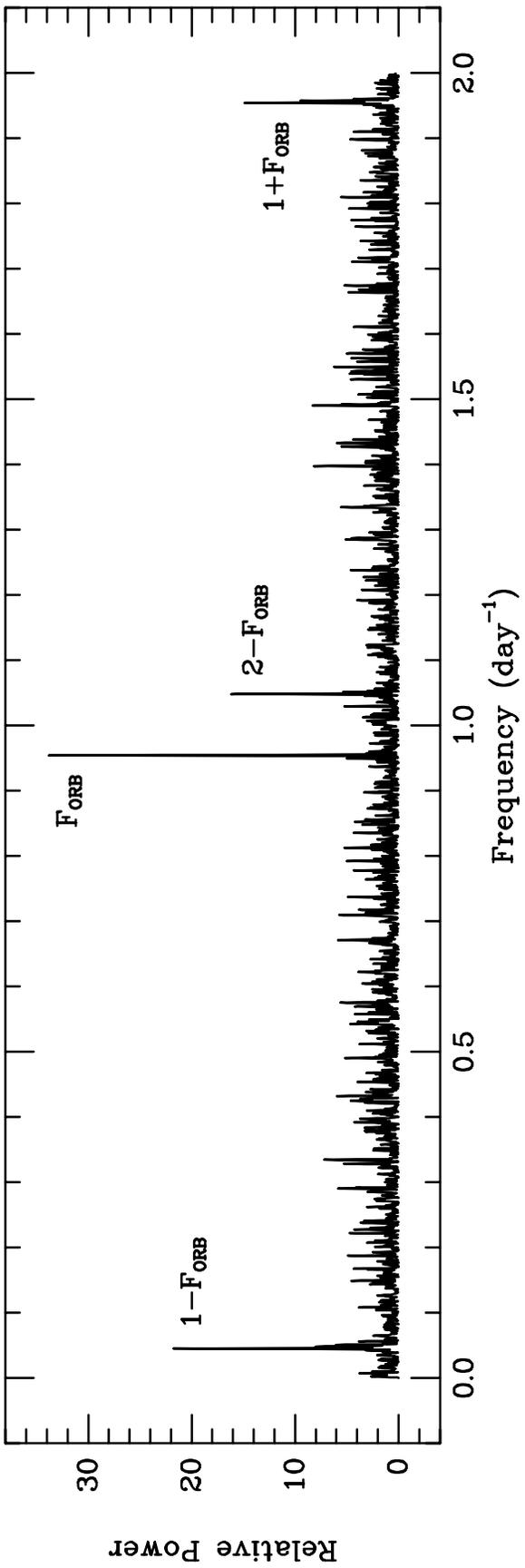
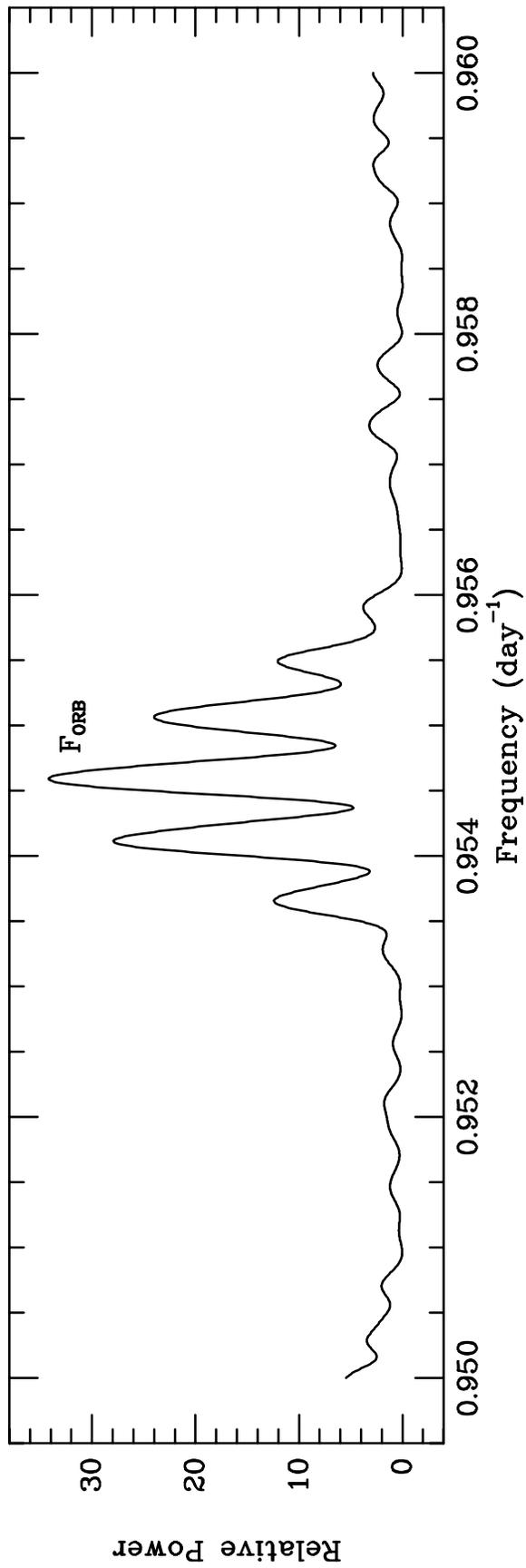
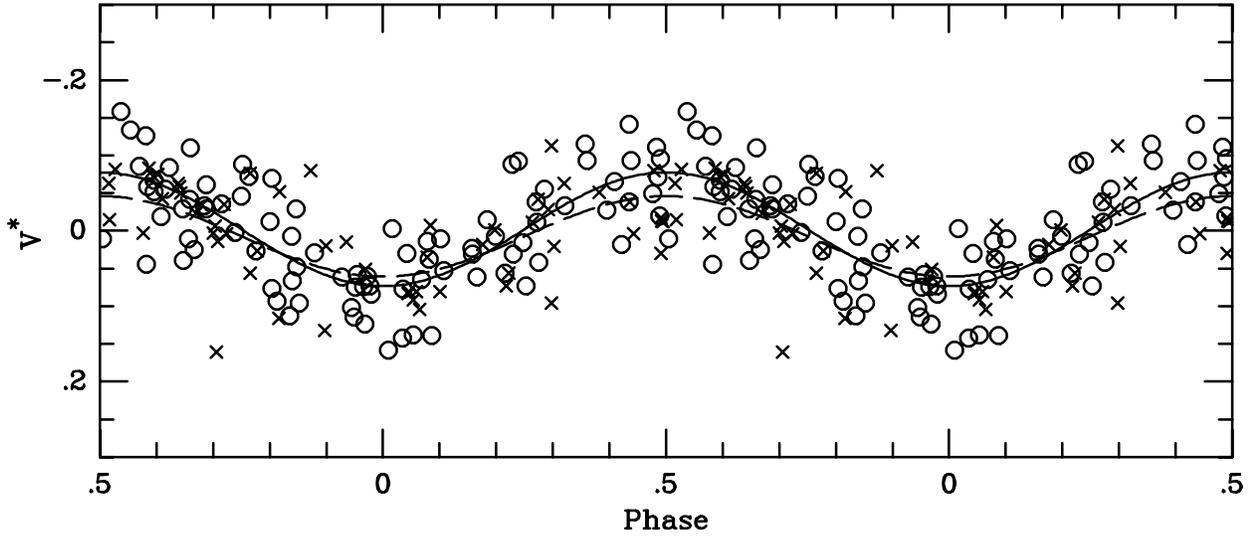
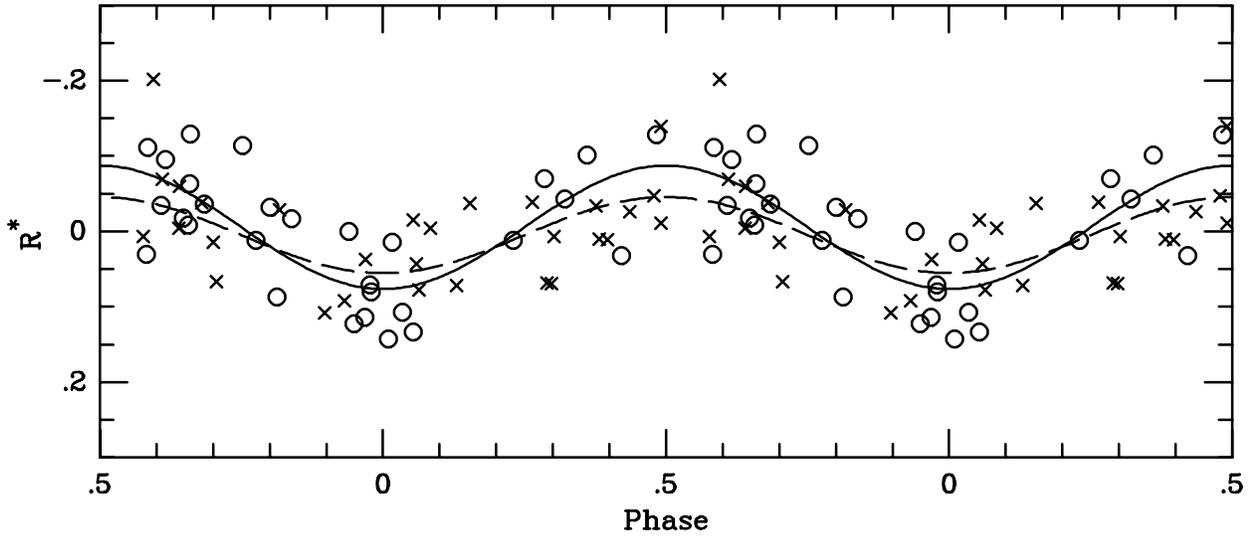
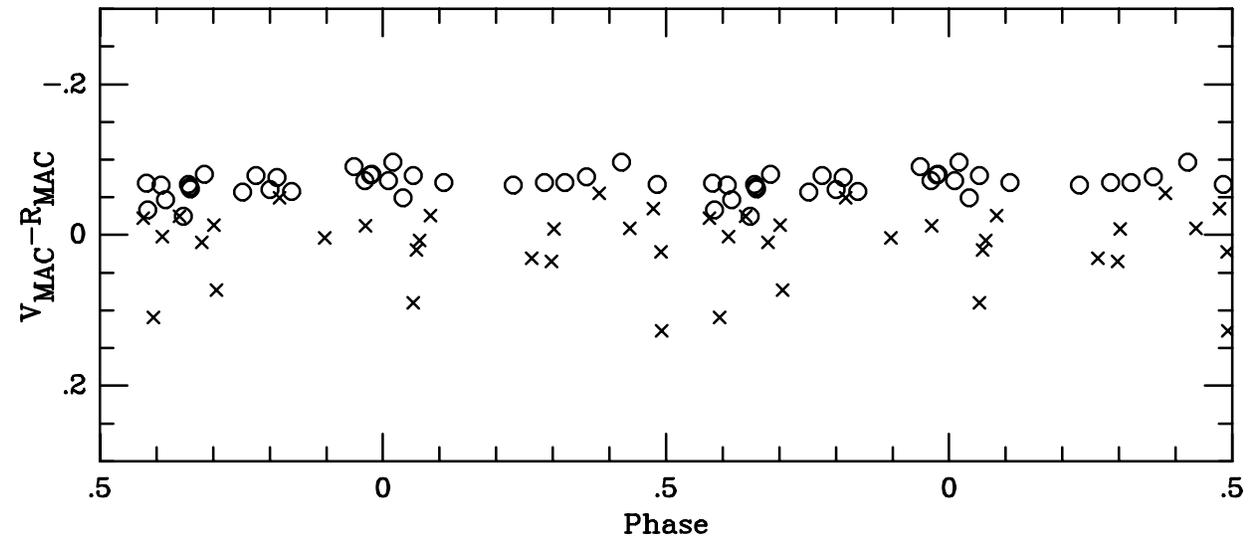
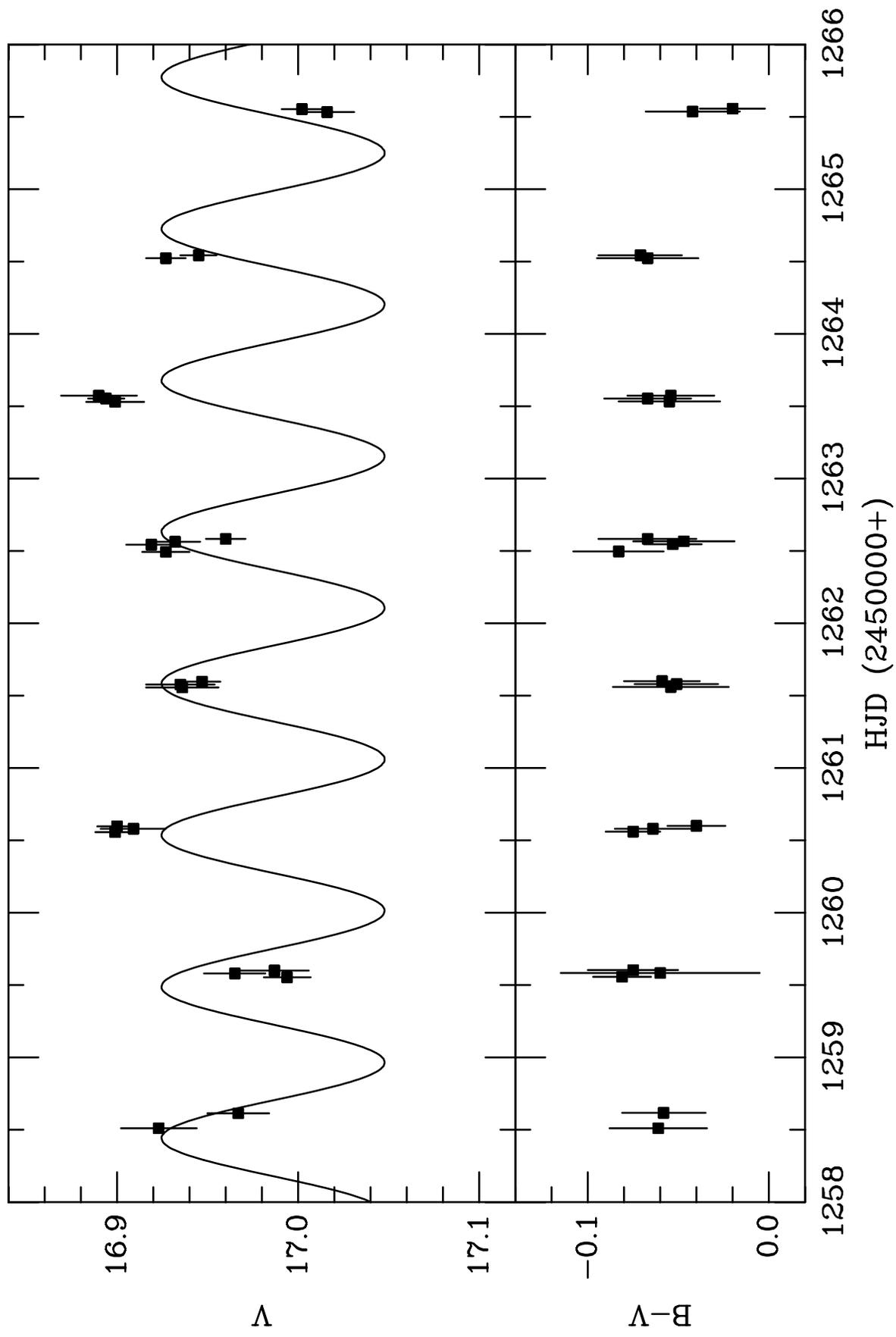
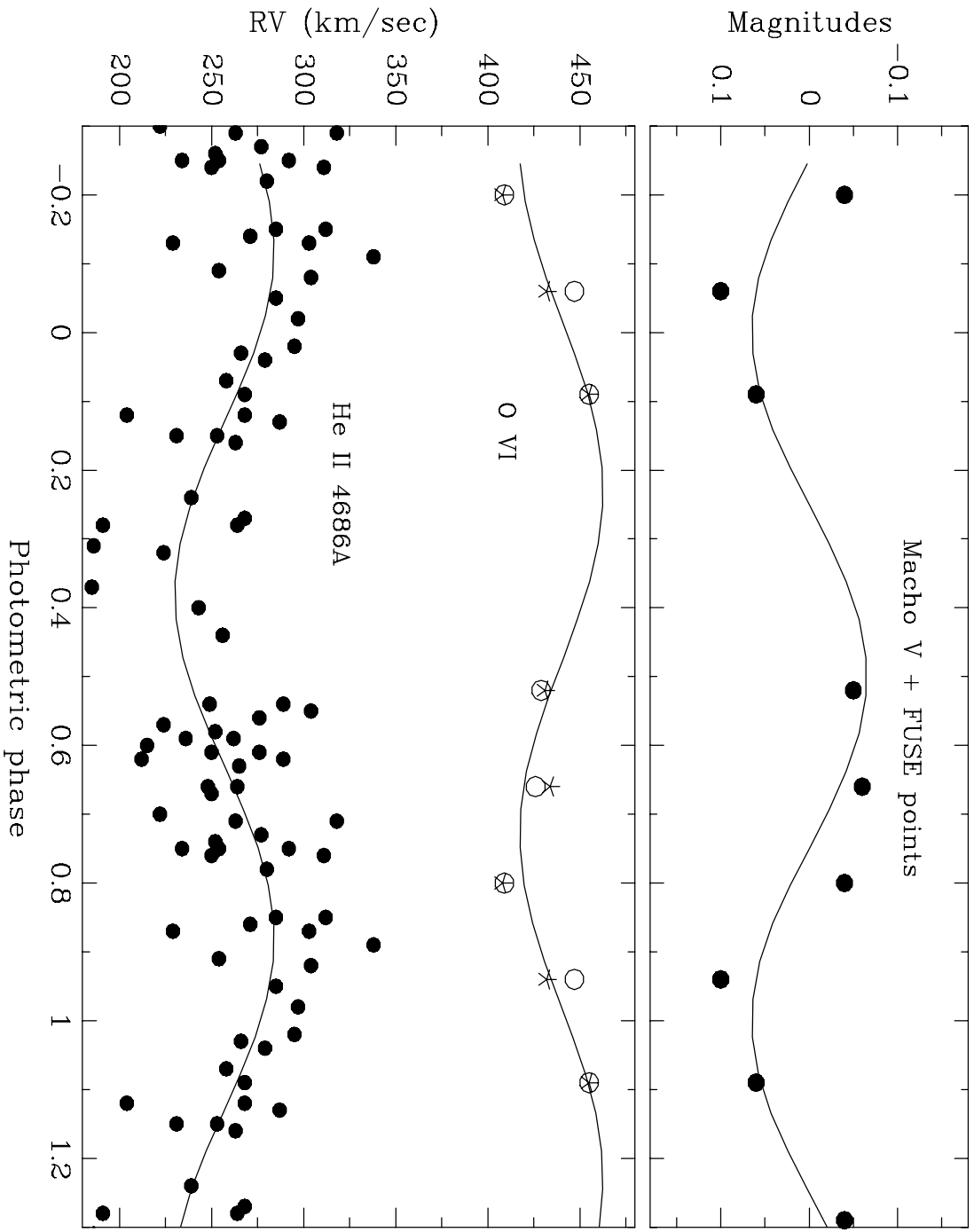
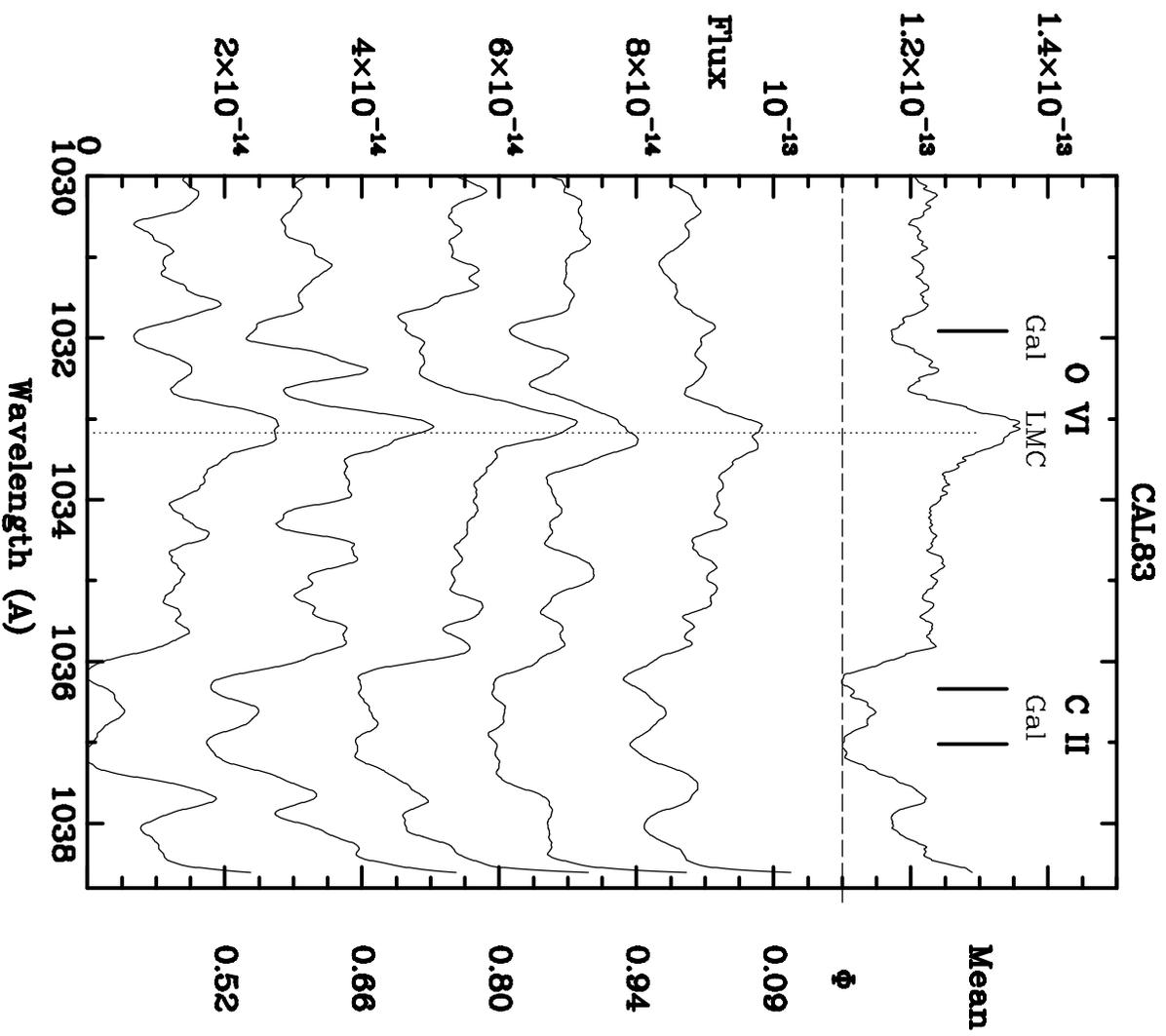
\end{document}